\documentclass[useAMS,usenatbib]{mn2e}
\usepackage{rotating}
\usepackage{graphicx}
\title[Classical Cepheids: Yet another version of the Baade--Becker--Wesselink method]{Classical Cepheids: Yet another version of the Baade--Becker--Wesselink method}
\author[A. S. Rastorguev, A. K. Dambis]{A. S. Rastorguev\thanks{E-mail:
rastor@sai.msu.ru.}, A. K. Dambis\\
Sternberg Astronomical Institute, Universitetskii pr. 13, Moscow, 119992 Russia\\}
\begin{document}
\date{Submitted to: Astrophysical Bulletin, 2011. Received: 2010}

\pagerange{\pageref{firstpage}--\pageref{lastpage}} \pubyear{2010}

\maketitle

\label{firstpage}

\begin{abstract}
We propose a new version of the Baade--Becker--Wesselink
technique, which allows one to independently determine the colour
excess and the intrinsic colour of a radially pulsating star, in
addition to its radius, luminosity, and distance. It is considered
to be a generalization of the Balona approach. The method also
allows the function $F(CI_0) = BC(CI_0) + 10 \times log (T_{eff}
(CI_0))$ for the class of pulsating stars considered to be
calibrated. We apply this technique to a number of classical
Cepheids with very accurate light and radial-velocity curves and
with bona fide membership in open clusters (SZ~Tau, CF~Cas, U~Sgr,
DL~Cas, GY~Sge), and find the results to agree well with the
reddening estimates of the host open clusters. The new technique
can also be applied to other pulsating variables, e.g. RR~Lyraes.
\end{abstract}


\section{Introduction}

Classical Cepheids are the key standard candles, which are used to set the zero point of the
extragalactic distance scale \citep{F01} and also serve as young-population tracers
of great importance \citep{BM}. They owe their popularity to their high luminosities and
photometric variability (which make them easy to identify and observe even
at large distances) and the fact that the luminosities, intrinsic colours, and ages of these stars
are closely related to such an easy to determine quantity as the variability period.

It would be best to calibrate the Cepheid period-luminosity (PL),
period-colour (PC), and period-luminosity-colour relations via
distances based on trigonometric parallaxes, however, the most
precisely measured parallaxes of even the nearest Cepeheids remain
insufficiently accurate and, more importantly, they may be fraught
with so far uncovered systematic errors. Here the
Baade--Becker--Wesselink method \citep{Baa26, Beck40, Wes46} comes
in handy, because it allows the Cepheid distances (along with the
physical parameters of these stars) to be inferred, thereby
providing an independent check for the results based on geometric
methods (e.g., trigonometric and statistical parallax).

However, all the so far proposed versions of the
Baade--Becker--Wesselink method (surface brightness technique
\citep{BE76}, maximum-likelihood technique \citep{B77}) depend, in
one way or another, on the adopted reddening value. Both
techniques are based on the same astrophysical background but make
use somewhat different calibrations (limb-darkened surface
brightness parameter, bolometric correction -- effective
temperature pair) on the normal colour. Here we propose a
generalization of the \citet{B77} technique, which allows one to
independently determine not only the star's distance and physical
parameters, but also the amount of interstellar reddening, and
even calibrate the dependence of a linear combination of the
bolometric correction and effective temperature on intrinsic
colour.

\section[]{Theoretical background}

We now briefly outline the method. First, the bolometric
luminosity of a star at any time instant is given by the following
relation, which immediately follows from the Stefan--Boltzmann
law:
\begin{equation}
L/L_{\odot} = (R/R_{\odot})^2 \times (T/T_{\odot})^4.
\end{equation}
Here $L$, $R$, and $T$ are the star's bolometric luminosity,
radius, and effective temperature, respectively, and the $\odot$
subscript denotes the corresponding solar values. Given that the
bolometric absolute magnitude $M_{bol}$ is related to bolometric
luminosity as
\begin{equation}
M_{bol} = M_{{bol}\odot} - 2.5 \times log (L/L_{\odot}),
\end{equation}
we can simply derive from Eq. (1):
\begin{equation}
M_{bol} - M_{{bol}\odot} = -5 \times log (R/R_{\odot})- 10 \times
log (T/T_{\odot})
\end{equation}
Now, the bolometric absolute magnitude $M_{bol}$ can be written in terms of the absolute
magnitude $M$ in some photometric band and the corresponding bolometric correction
$BC$:
\begin{equation}
M_{bol} = M + BC,
\end{equation}
and the absolute magnitude $M$ can be written as:
\begin{equation}
M = m - A - 5 \times log (d/10~pc).
\end{equation}
Here $m$, $A$, and $d$ are the star's apparent magnitude and interstellar extinction in the
corresponding photometric band, respectively, and $d$ is the heliocentric distance
of the star in pc. We can therefore rewrite Eq. (3) as follows:
$$
m=A+5\times log (d/10~pc)+M_{{bol}\odot}+10\times log(T_{\odot})
$$
\begin{equation}
-5\times log (R/R_{\odot}) -BC -10\times log (T).
\end{equation}
Let us introduce the function $F(CI_0) = BC + 10\times \log (T)$,
the apparent distance modulus $(m-M)_{app} = A + 5\times log
(d/10~pc)$, and rewrite Eq. (7) as the light curve model:
\begin{equation}
m = Y - 5\times log (R/R_{\odot}) - F.
\end{equation}
where constant
$$
Y = (m-M)_{app} + M_{{bol}\odot}+10\times log(T_{\odot}).
$$

We now recall that interstellar extinction $A$ can be determined
from the colour excess $CE$ as $A = R_{\lambda} \times CE$, where
$R_{\lambda}$ is the total-to-selective extinction ratio for the
passband-colour pair considered, whereas $M_{{bol}\odot}$,
$R_{\odot}$, and $T_{\odot}$ are rather precisely known
quantities. The quantity $F(CI_0) = BC + 10\times log (T)$ is a
function of intrinsic colour index $CI_0 = CI - CE$. \citet{B77}
used a very crude approximation for the effective temperature and
bolometric correction, reducing the right-hand of the light curve
model (Eq. 7) to the linear function of the observed colour, with
the coefficients containing the colour excess in a latent form. It
should be noted that \citet{Sach98, Sach02} used non-linear
approximation in Eq.~(7) to calculate Cepheid radii.

The key point of our approach is that the values of function $F$ are
computed from the already available calibrations of bolometric
correction $BC(CI_0)$ and effective temperature $T(CI_0)$
~\citep{F96, BCP98, AAMR99, SF00, RM05, BFCM07, GHB09}. These
calibrations are expressed as high-order power series in the
intrinsic colour:
\begin{equation}\label{Ffun}
F(CI_0) = a_0 + \sum^N_{k=1}a_k \cdot CI_0^k,
\end{equation}
with known $\{a_k\}$ and $N$ amounting to $7$; in some cases the
decomposition also includes the metallicity ($[Fe/H]$) and/or
gravity ($log~g$) terms.

As for the star's radius $R$, its current value can be determined
by integrating the star's radial-velocity curve over time ($dt = (P/ 2\pi)\cdot d\varphi$):
\begin{equation}\label{bw}
R(t)-R_0 =-pf \cdot \int^{\varphi}_{\varphi_0}{(V_r(t) -
V_{\gamma})\cdot (P / 2\pi) \cdot d\varphi},
\end{equation}
where $R_0$ is the radius value at the phase $\varphi_0$ (we use
mean radius, $<R>=(R_{min}+R_{max})/2$ ); $V_{\gamma}$, the
systemic radial velocity; $\varphi$, the current phase of the radial
velocity curve; $P$, the star's pulsation period, and $pf$ is the
projection factor that accounts for the difference between the
pulsation and radial velocities. Given the observables (light
curve -- apparent magnitudes $m$, colour curve -- apparent colour
indices $CI$, and radial velocity curve -- $V_r$) and known
quantities for the Sun, we end up with the following unknowns:
distance $d$, mean radius $<R>$, and colour excess $CE$, which can be
simply found by the least-squares or maximum likelihood technique.

In the case of Cepheids with large amplitudes of light and colour curves
($\Delta CI \geq 0.4^m$) it is also possible to apply a more
general technique by setting the expansion coefficients $\{a_k\}$
in Eq.~(8) free and treating them as unknowns. We expanded the
function $F = BC + 10\times log (T)$ in Eq.~(7) into a power
series about the intrinsic colour index $CI_0^{st}$ of a
well-studied ``standard'' star (e.g., $\alpha$ Per or some other
bright star) with accurately known $T^{st}$:
\begin{equation}\label{Fgen1}
F = BC^{st} + 10\times log (T^{st}) + \sum^N_{k=1}a_k \cdot
(CI-CE-CI_0^{st})^k
\end{equation}

The best fit to the light curve is provided with the optimal expansion
order $N \simeq 5 - 9$. We use this modification to calculate the
physical parameters and reddening $CE$ of the Cepheid, as well as
the calibration $F(CI_0) = BC(CI_0) + 10 \times log (T_{eff}
(CI_0))$ for the given metallicity $[Fe/H]$.

\section{Observational data, constants, and calibrations}

Our sources of data include Berdnikov's extensive multicolor
photoelectric and CCD photometry of classical Cepheids~\citep{B95,
B08} and very accurate radial-velocity measurements of 165
northern Cepheids~\citep{G92, G96, G98, G02} taken in 1987-2009
(about 10500 individual observations) with a CORAVEL-type
spectrometer \citep{Tok87}. These data sets are nearly
synchronous, to prevent any systematic errors in the computed
radii and other parameters due to the evolutionary period changes
resulting in phase shifts between light, colour and radial
velocity variations \citep{Sach98}. We adopt $T_{\odot}$ = 5777 K,
$M_{bol \odot} = +4.76^m$ \citep{G05}. We proceeded from $(V,
B-V)$ data and found the best solutions for the $V$-band light
curve and $(B-V)$ color curve to be those computed using the
$F((B-V)_0)$ function based on two calibrations \citep{F96, BCP98}
of similar slope (see Fig.~2~e); the poorer results obtained using
the other cited calibrations can be explained by the fact that the
latter involved insufficient number of supergiant stars.

\section{The projection factor}

There is yet no consensus concerning the projection factor (PF)
value to be used for Cepheid variables ~\citep{Nar04, Gr07, Nar07,
Nar09}. Different authors use constant values ranging from 1.27 to
1.5, as well as PFs depending on the pulsation period and other
parameters. Different approaches lead to small systematic
differences in the inferred Cepheid parameters, first of all, in
the radii. Based on geometrical considerations, ~\citet{Ras10}
derived phase-dependent PFs as simple three-parametric analytic
expressions depending on the pulsation velocity, limb darkening
coefficient, and spectral line broadening, adjusted to CORAVEL
radial velocities of Cepheids. We suspect that the period
dependence reflects mainly the dependence of the PF on limb
darkening. To compare our results with other calculations, we
finally adopted a moderate dependence of PF on the period
advocated by~\citet{Nar07}:
\begin{equation}\label{Fgen2}
p = (-0.064 \pm 0.020) \times log (P, days) + (1.376 \pm 0.023),
\end{equation}
though we repeated all calculations with other variants of PF
dependence on the period and  pulsation phase to assure the
stability of the calculated colour excess.

\section{Preliminary results}

To test the new method, we used the maximum likelihood technique
to solve Eq.~(7) for the $V$-band light curve and $B-V$ colour
curve for several classical Cepheids residing in young open
clusters: SZ~Tau (NGC~1647), CF~Cas (NGC~7790), U~Sgr (IC~4725),
DL~Cas (NGC~129), GY~Sge (anonymous OB-association~\citep{Forb82})
as well as for approximately 30 field Cepheids from our sample. We
found two $log (T_{eff})$ calibrations -- those of~\citet{F96} and
~\citet{BCP98} -- combined with the $BC(V)$ calibration as a
function of normal colour $(B-V)_0$ proposed by ~\citet{F96} -- to
yield the best fit to the observed $V$-band light curve via
Eq.~(7). A weak sensitivity of calculated reddening, $E_{B-V}$, to the adopted
PF value (constant or period/phase--dependent) and to the derived
$<R>$ value can be explained by very strong dependence of the
light curve's amplitude on the effective temperature, $\sim 10
\times log(T)$, and, as a consequence, on the dereddened colour.

Though the internal errors of the reddening $E_{B-V}$ seem to be
very small, the values determined using the two best calibrations,
\citet{F96} and \citet{BCP98}, may differ by as much as $0.03 -
0.05^m$, due to the systematic shift between these two calibrations
(Fig. 2~e). Table~\ref{tab1} summarizes the inferred parameters
for the cluster Cepheids studied. Fig.~1 shows the observed and
smoothed data and the final fit to the $V$-band light curve for
U~Sgr Cepheid. Our reddenings seem to agree well with the
corresponding WEBDA values, particularly if we remember that the
errors of the adopted cluster reddening estimates
are as high as $\pm 0.05^m$. Our next step will
be to make use of the calibrations of $T_{eff}$ and $BC$ as a
function of red and infrared colours $(V-R, ~V-I, ~V-K)$ and to
compare derived reddening ratios with the conventional extinction
laws.

Note that the inferred radius and luminosity of SZ~Tau are too
large for its short period; this Cepheid probably pulsates in the
$1^{st}$ or even in the $2^{nd}$ overtone, as may be indirectly
evidenced by its low colour amplitude (about $0.15^m$).

Fig.~2 shows the observed data, the fit to the $V$-band light
curve, and the inferred calibration $F = 10\times log (T_{eff}) +
BC(V)$ vs $(B-V)_0$) calculated for TT~Aql Cepheid (as a
$5^{th}$-order expansion in the normal colour). The inferred
calibration is very close to that of \citet{F96}. We used
$\alpha$~Per as the ``standard'' star, with $T^{st} \approx
(6240\pm20)~K$, $[Fe/H]\approx -0.28\pm0.06$ \citep{Lee06},
$(B-V)^{st}\approx 0.48^m$ and $E_{B-V}\approx 0.09^m$ (WEBDA, for
$\alpha$~Per cluster). To take into account the effect of
metallicity on the zero-point $F(CI_0)^{st}$, we estimated the
gradient $dF(CI_0)^{st} / d[Fe/H]\approx +0.24$ from the
calibrations by \citet{AAMR99, SF00, GHB09}. For TT~Aql,
$E_{B-V}\approx (0.65\pm0.03)^m$. In some cases (with large
amplitude of color variation) the ``free'' calibration (Eq.~10)
can markedly improve the model fit to the observed light curve of
the Cepheid variable. Fig. 2~f shows the example of calibrations
of the $F$ functions derived from nine Cepheids with different
metallicity and surface gravity values. Temperature difference at
$T_{eff} \sim 6600 - 5100~K$ is amounted to $3 - 5\%$.

When applied to an extensive sample of Cepheid
variables with homogeneous photometric data and detailed radial
velocity curves, the new method is expected to give a completely independent scale
of reddenings, a new Period - Colour - Luminosity relation, and a new
distance scale for the Milky-Way Cepheids.

\section{Acknowledgements}
We grateful to M.V.~Zabolotskikh for her assistance in data
preparation and to L.N.~Berdnikov, Yu.N.~Efremov, M.E.~Sachkov,
V.E.~Panchuk and A.B.~Fokin for comments and helpful discussions.
This research has made use of the WEBDA database operated at the
Institute for Astronomy of the University of Vienna. Our work is
supported by the Russian Foundation for Basic Research (projects
nos.~08-02-00738-a, 07-02-00380-a, and 06-02-16077-a).

\begin{table*}
  \centering
  \begin{minipage}{140mm}
  \caption{Physical parameters, distances, and interstellar reddening values for
the cluster Cepheids analysed using the new version of the BW
method. Reddening values from WEBDA data base
(http://www.univie.ac.at/webda/) are also shown for comparison;
asterisk: $E_{B-V}$ adopted from \citet{Forb82}. Distances are
calculated using $R_V=3.3$.}\label{tab1}
\begin{tabular}{l l l l l l l l}
\hline
  Star & Cluster & Period (d) & Distance (pc) & $E_{B-V}$ & $E_{B-V}$(WEBDA) & $<R>/R_{\odot}$ & $M_V$ \\
  \hline
SZ~Tau  & NGC~1647  &  ~3.149  & ~796$\pm$90   & 0.40$\pm$0.02  & 0.370       & ~57.0$\pm$7.0 & -4.32$\pm$0.25 \\
CF~Cas  & NGC~7790  &  ~4.875  &  3585$\pm$87  & 0.54$\pm$0.02  & 0.531       & ~46.7$\pm$0.9 & -3.41$\pm$0.05 \\
~~U~Sgr & IC~4725   &  ~6.745  & ~613$\pm$25   & 0.50$\pm$0.03  & 0.475       & ~54.2$\pm$1.8 & -3.90$\pm$0.08 \\
DL~Cas  & NGC~129   &  ~8.001  &  2067$\pm$58  & 0.47$\pm$0.05  & 0.548       & ~69.3$\pm$1.6 & -4.12$\pm$0.06 \\
GY~Sge  & Anon OB   &  51.78   &  2136$\pm$163 & 1.44$\pm$0.05  & 1.29$\pm$0.06~(*) &  208$\pm$11  & -6.27$\pm$0.15 \\
\hline
\end{tabular}
\end{minipage}
\end{table*}

\begin{figure*}
(a){\label{fig:edge-a1}\includegraphics[angle=0,
width=6cm]{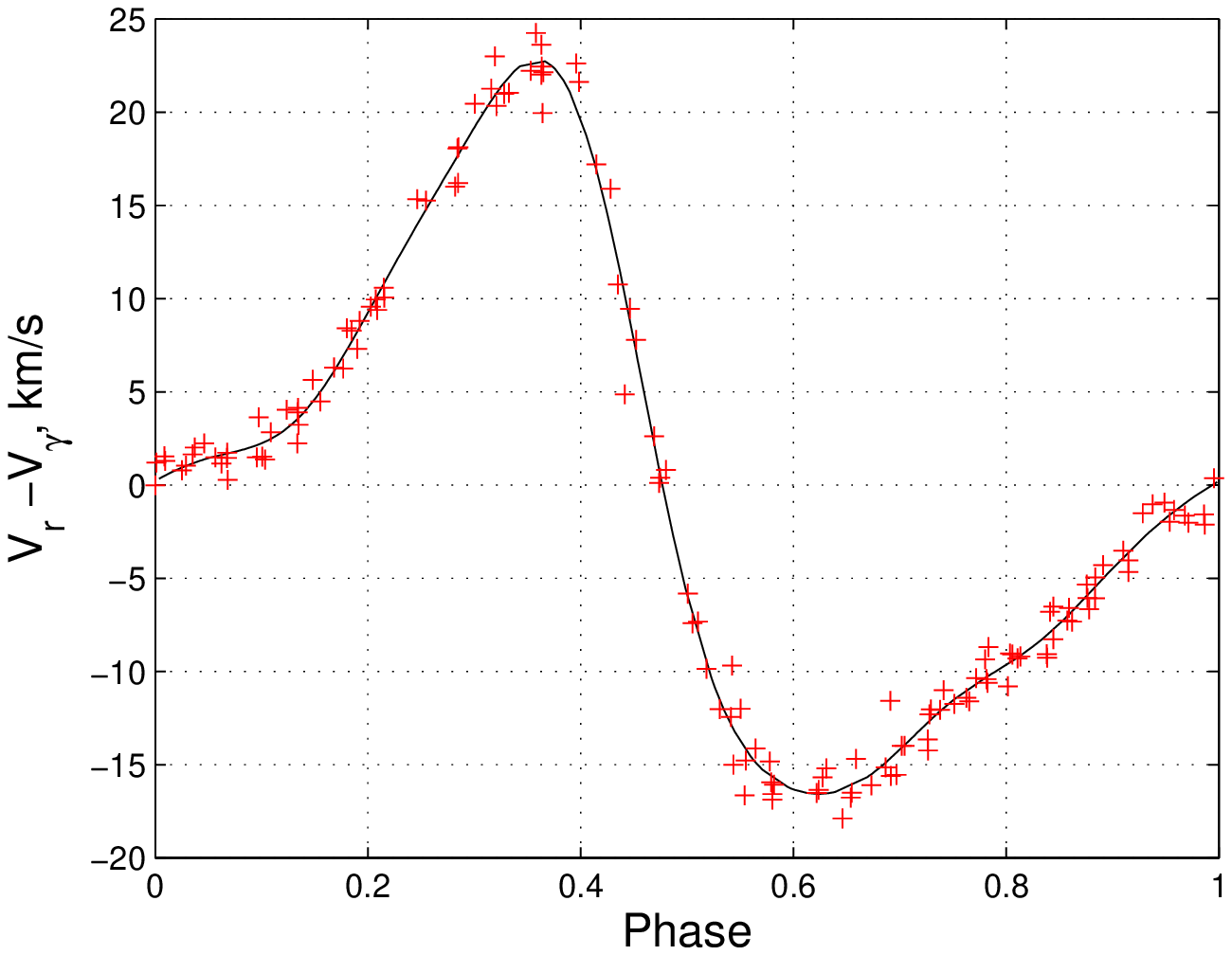}}
(b){\label{fig:edge-b1}\includegraphics[angle=0,
width=6cm]{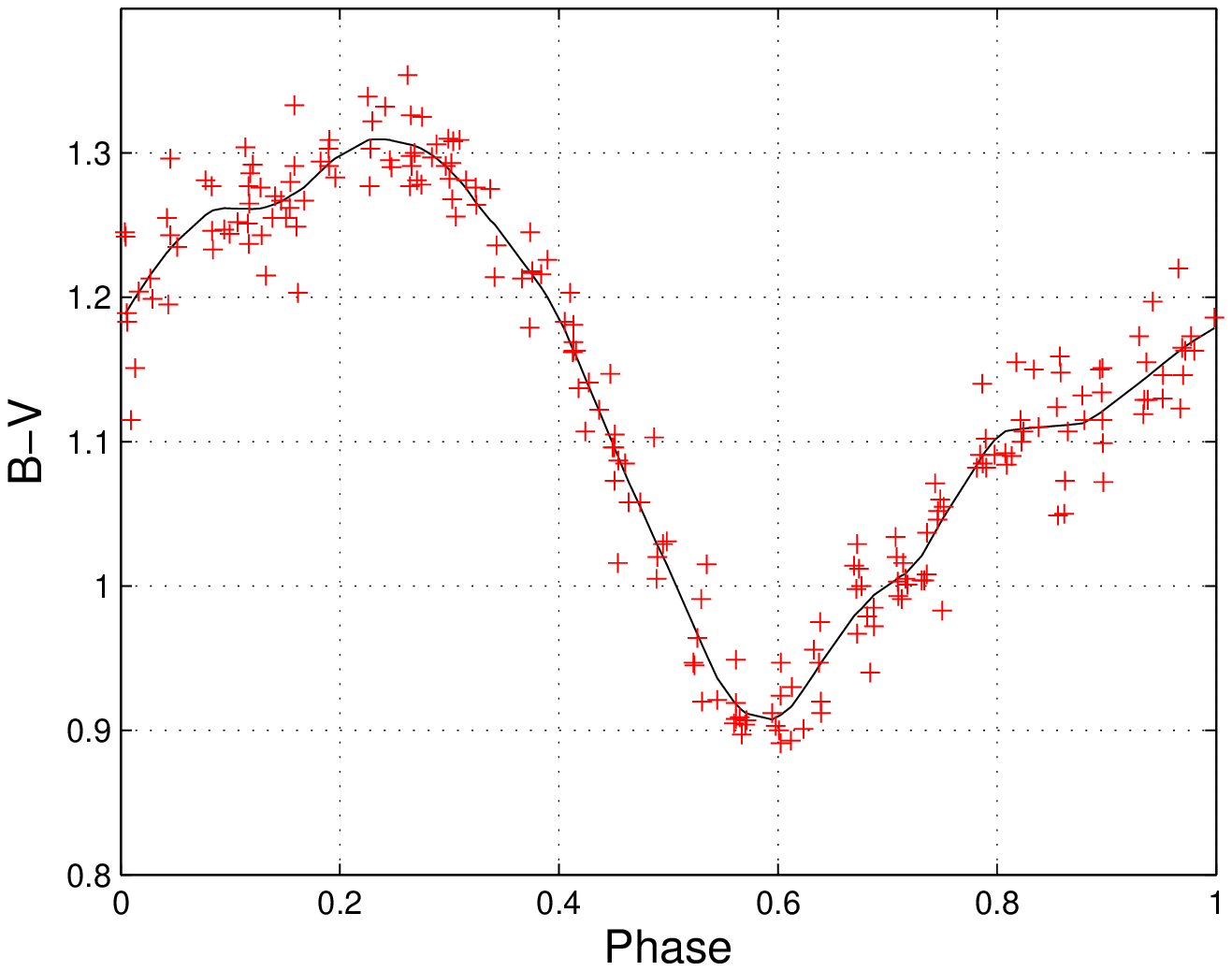}}
 \\
(c){\label{fig:edge-c1}\includegraphics[angle=0,
width=6cm]{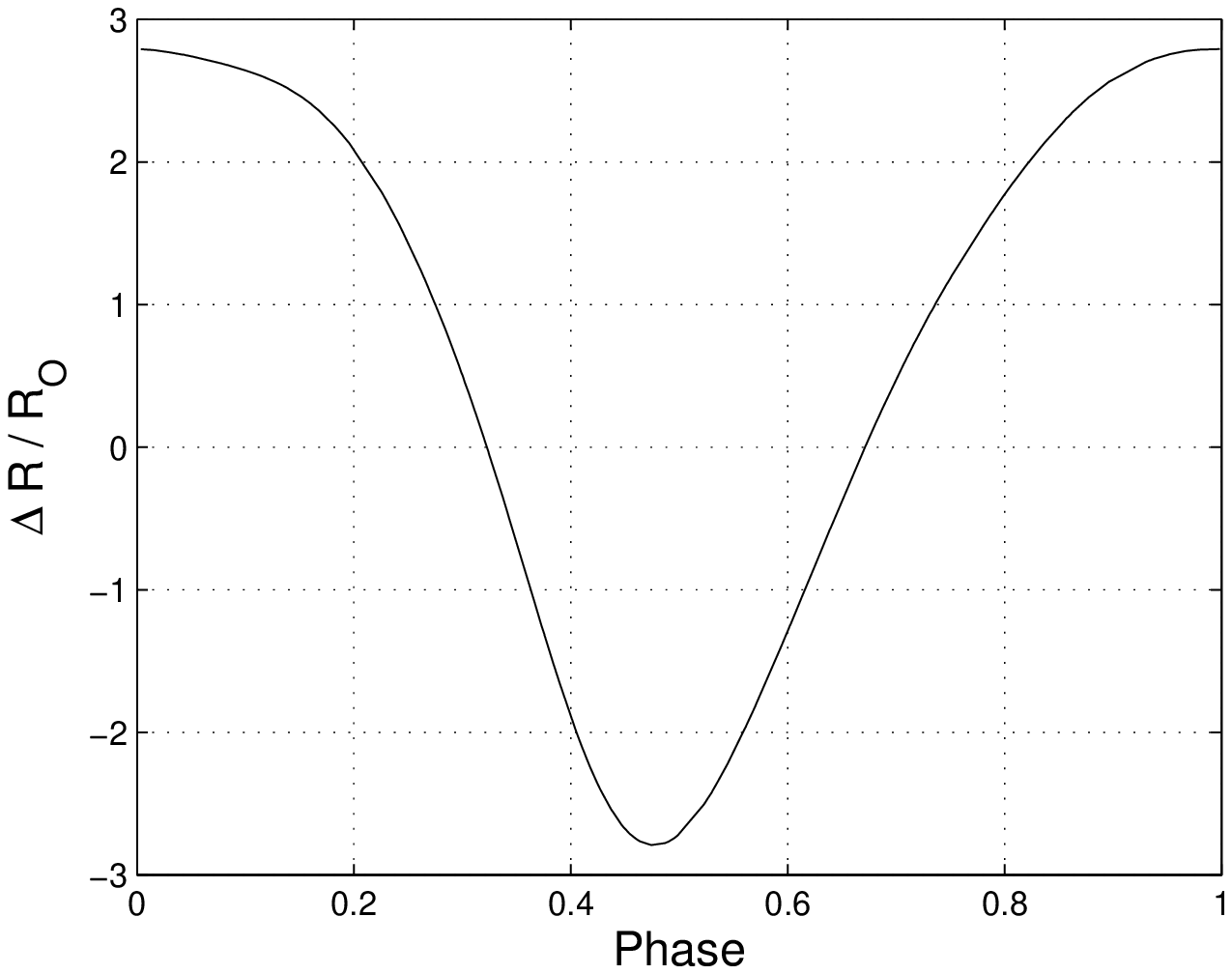}}
(d){\label{fig:edge-d1}\includegraphics[angle=0,
width=6cm]{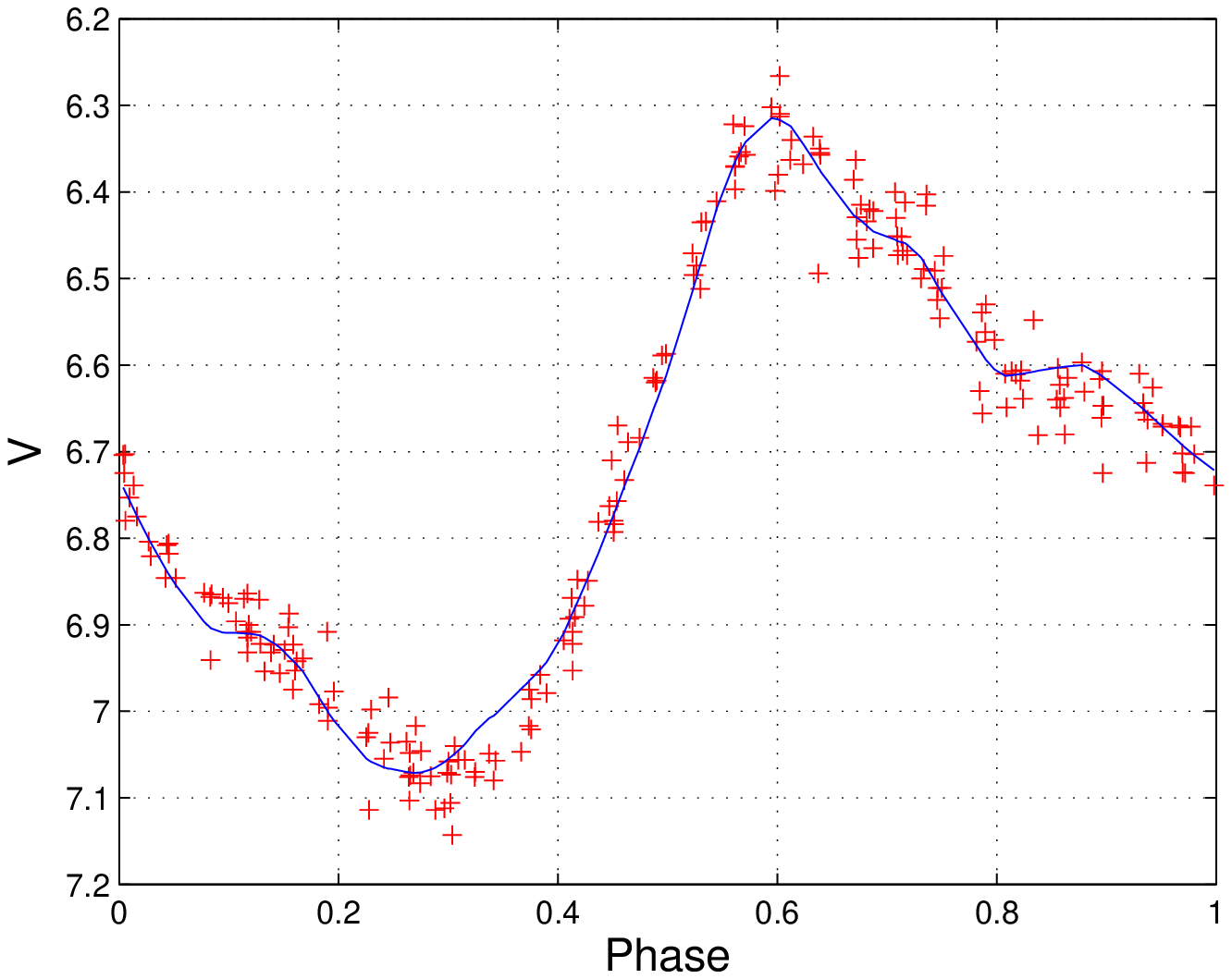}}
\\
\caption{Panel (a): Observed and fitted radial-velocity curve of
U~Sgr. Standard deviation $\sigma_{Vr} = 1.1~km/s$. Panel (b):
Observed and smoothed colour curve. Panel (c): Radius variation
with phase. Panel (d): Observed and fitted light curve.}
 \label{fig1}
\end{figure*}

\begin{figure*}
(a){\label{fig:edge-a2}\includegraphics[angle=0,
width=6cm]{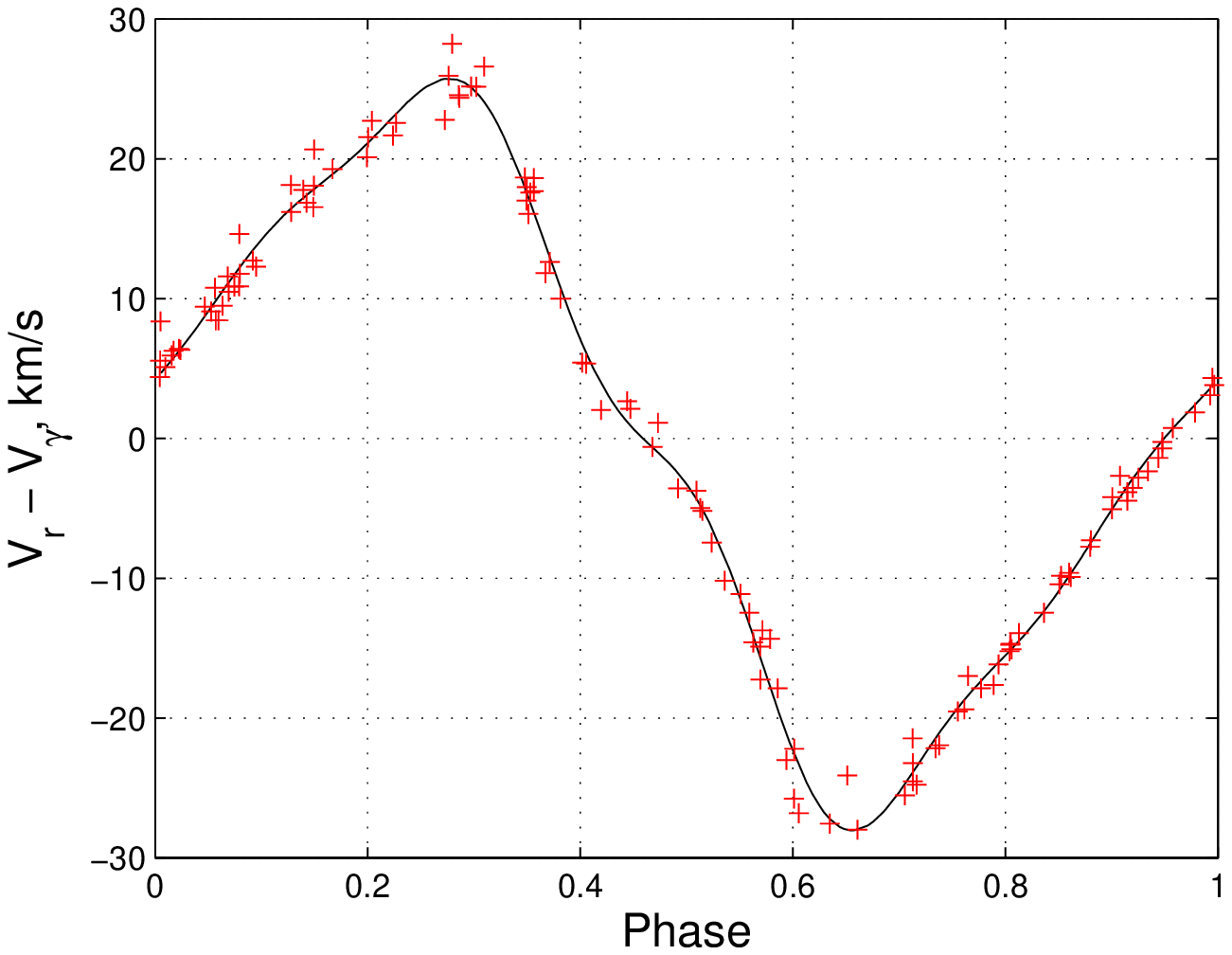}}
(b){\label{fig:edge-b2}\includegraphics[angle=0,
width=6cm]{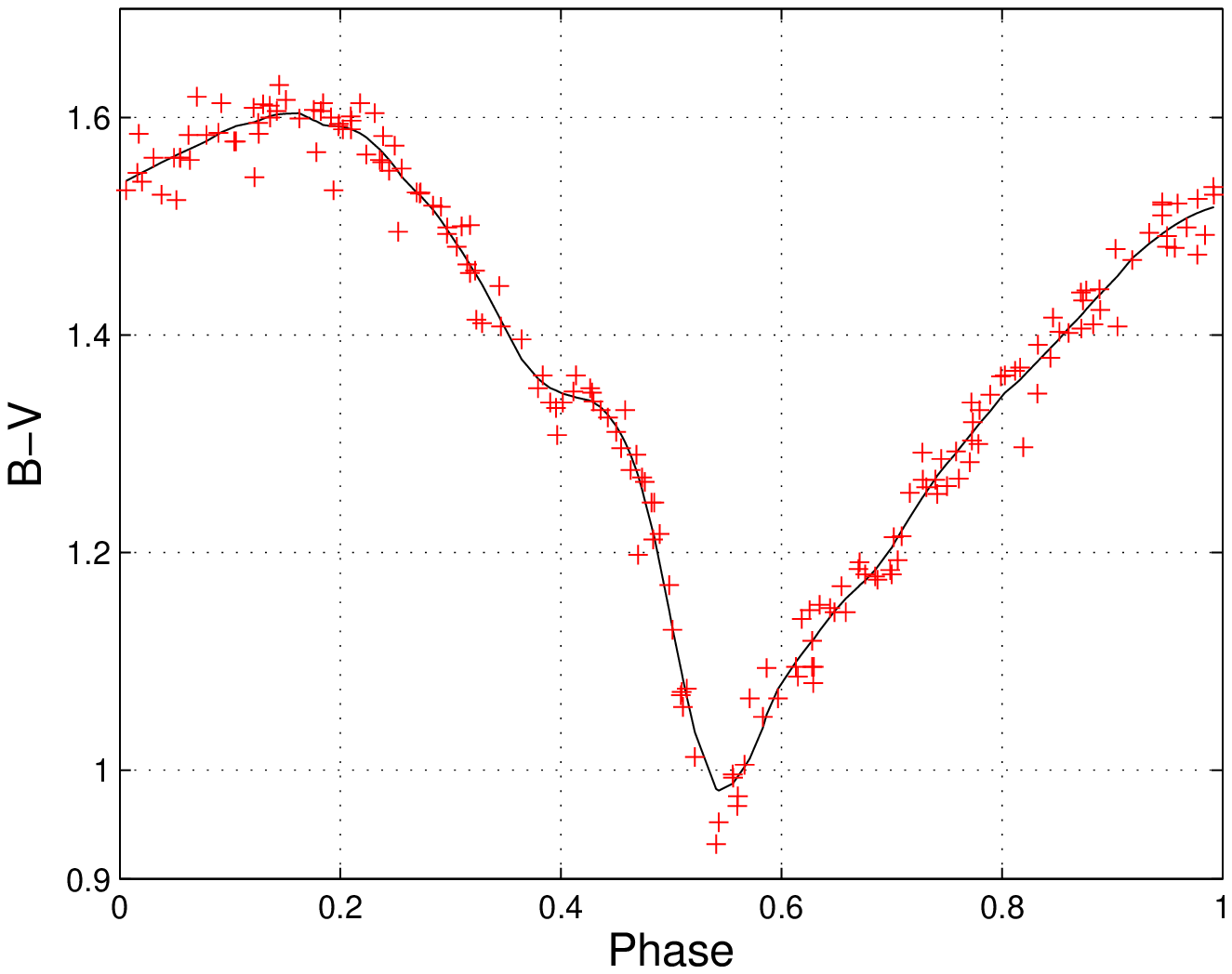}}
 \\
(c){\label{fig:edge-c2}\includegraphics[angle=0,
width=6cm]{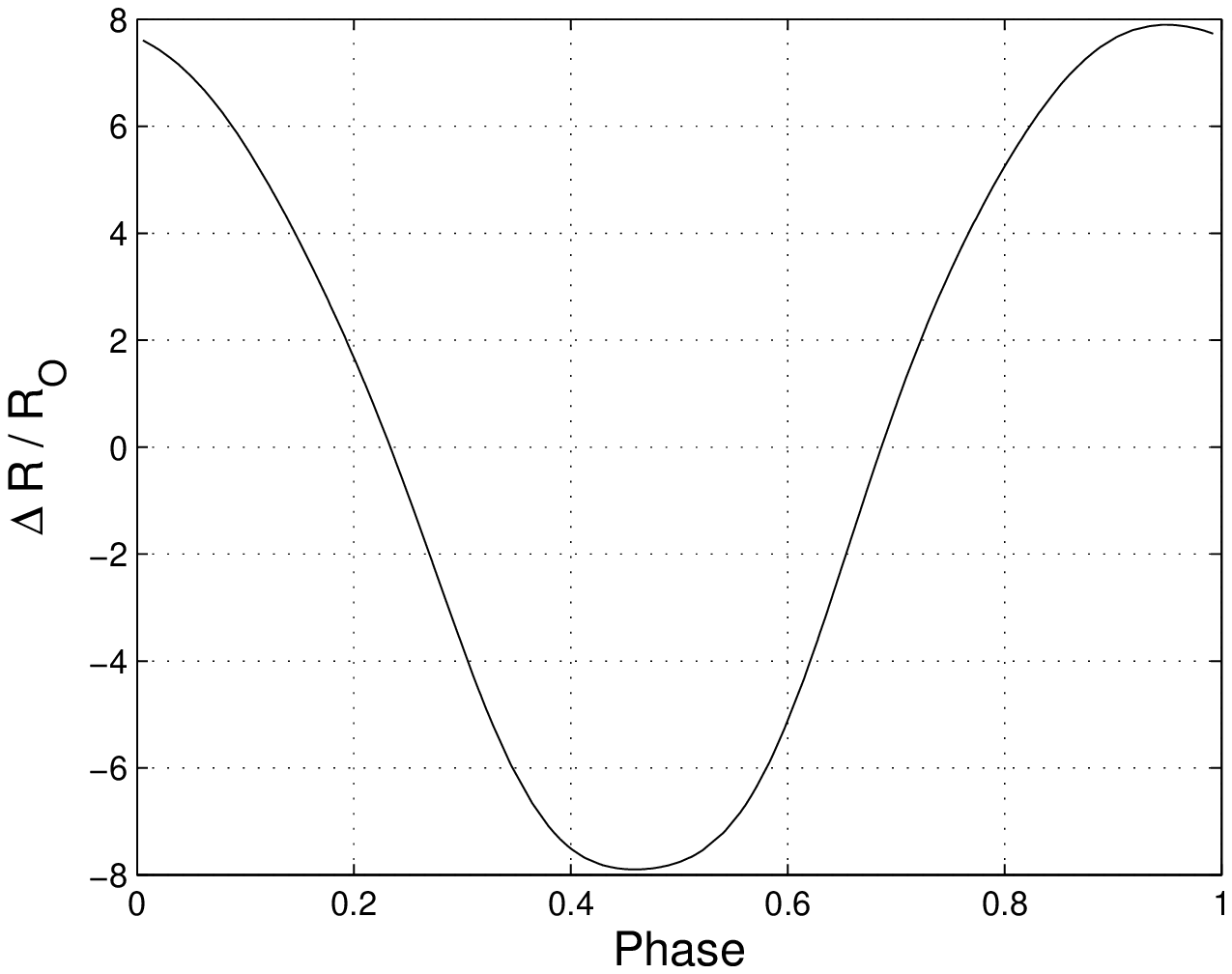}}
(d){\label{fig:edge-d2}\includegraphics[angle=0,
width=6cm]{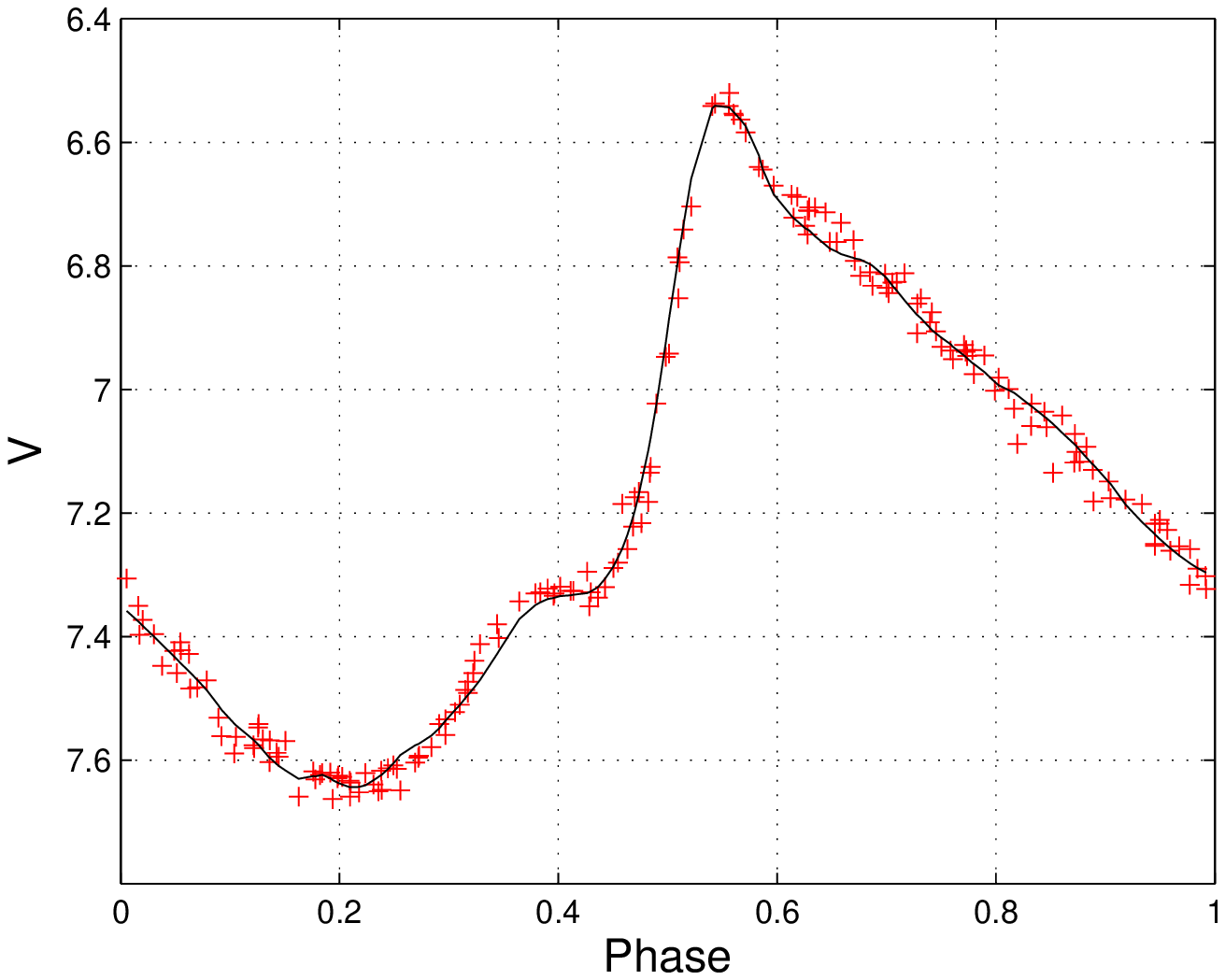}}
\\
(e){\label{fig:edge-a}\includegraphics[angle=0,
width=6cm]{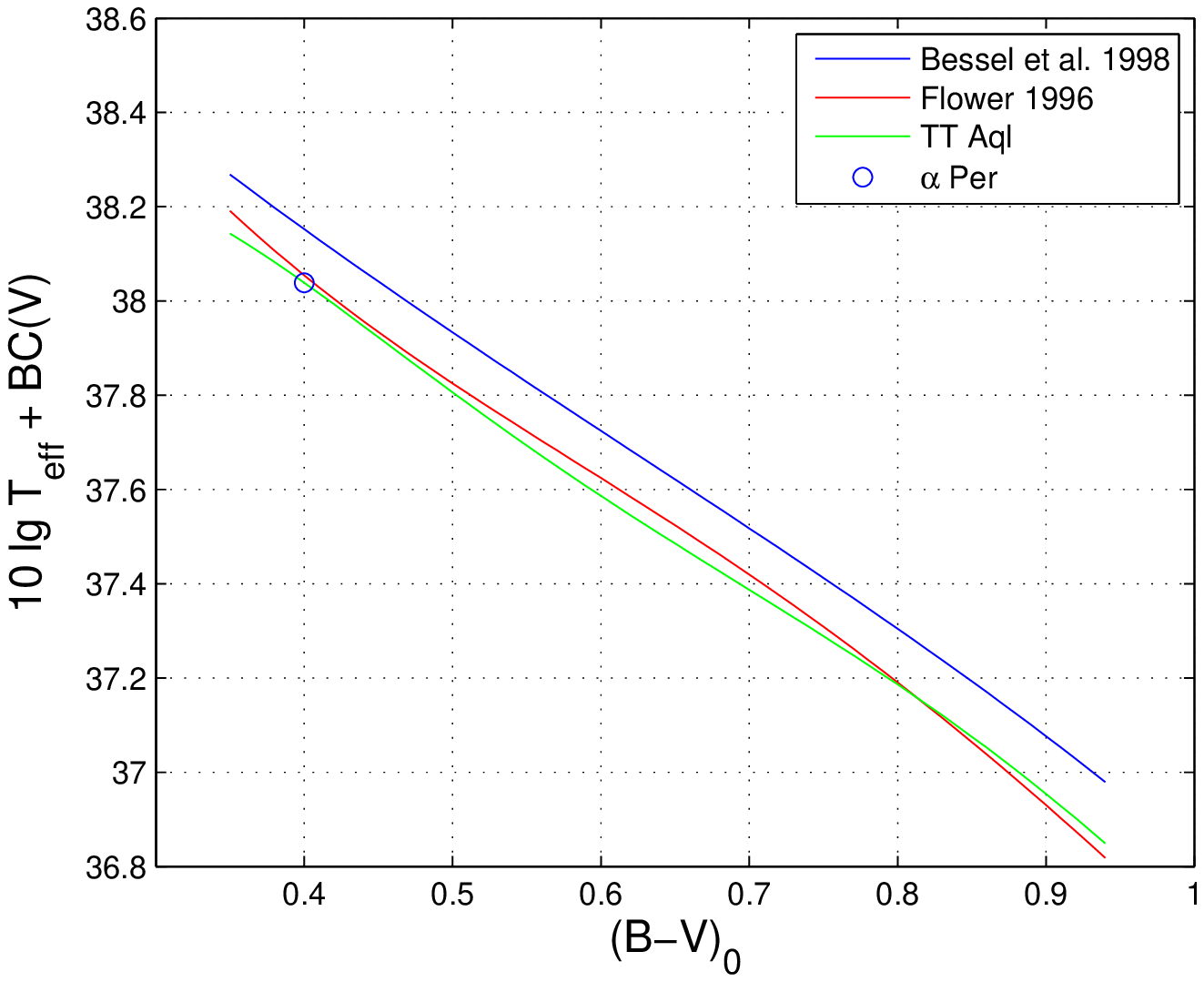}}
(f){\label{fig:edge-aa}\includegraphics[angle=0,
width=6cm]{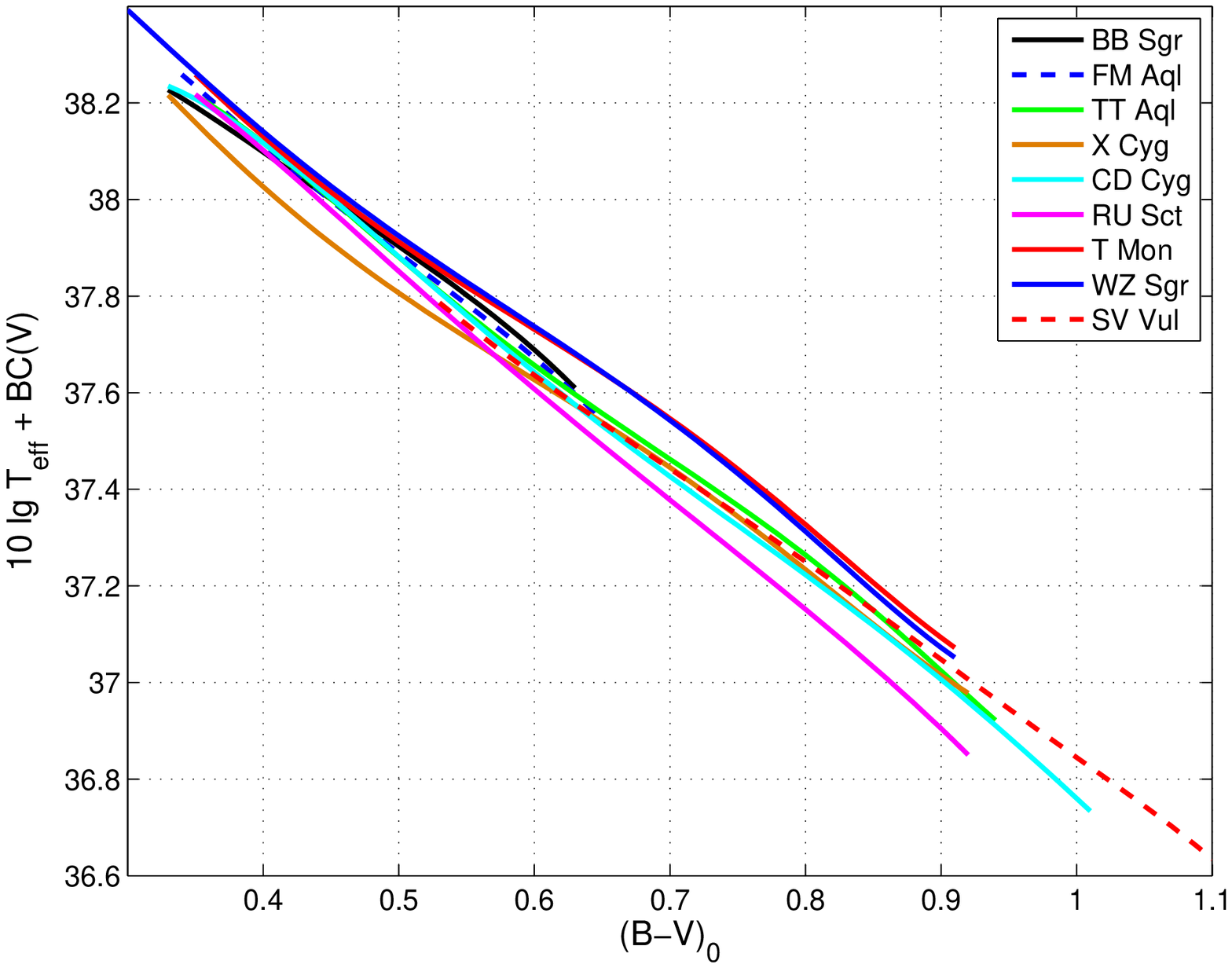}}
\\
\caption{Panel (a): Observed and fitted radial-velocity curve of
TT~Aql. Standard deviation $\sigma_{Vr} = 1.3~km/s$. Panel (b):
Observed and smoothed colour curve. Panel (c): Radius variation
with phase. Panel (d): Observed and fitted light curve. Panel (e):
Calculated calibration for TT~Aql (function $F = 10\times log
(T_{eff}) + BC(V) ~ vs ~ (B-V)_0$)) and calibrations by
\citet{F96} and \citet{BCP98}. Also shown is the position of the
standard star $\alpha$~Per corrected for metallicity difference.
Panel (f): Calculated calibration (function $F = 10\times log
(T_{eff}) + BC(V) ~ vs ~ (B-V)_0$)) for 9 Cepheids with large
amplitudes of the colour curves and different metallicities.}
\label{fig2}
\end{figure*}

\end{document}